\documentclass[prb,twocolumn,aps,floatfix,superscriptaddress]{revtex4-1}

\usepackage {graphicx}
\usepackage [centertags]{amsmath}
\usepackage {amsfonts}
\usepackage {amssymb}
\usepackage {mathrsfs}
\usepackage {bm}

\usepackage {epstopdf}

\newcommand {\be}{\begin{eqnarray}}
\newcommand {\ee}{\end{eqnarray}}
\newcommand {\intp}{\int \frac{d^{d+1}p}{(2 \pi)^{d+1}}}
\newcommand {\intq}{\int \frac{d^{d+1}q}{(2 \pi)^{d+1}}}
\newcommand {\intk}{\int \frac{d^{d+1}k}{(2 \pi)^{d+1}}}

\begin{document}

\title{Quantum Criticality for Extended Nodes on a Bethe Lattice in the Large Connectivity Limit}



\author{James M. Murray}
\affiliation{Institute for Quantum Matter and Department of Physics and Astronomy, Johns Hopkins University, Baltimore, Maryland 21218, USA}
\author{Adrian \surname{Del Maestro}}
\affiliation{Institute for Quantum Matter and Department of Physics and Astronomy, Johns Hopkins University, Baltimore, Maryland 21218, USA}
\affiliation{Department of Physics, University of Vermont, Burlington, Vermont 05405, USA}
\author{Zlatko Te\v{s}anovi\'{c}}
\affiliation{Institute for Quantum Matter and Department of Physics and Astronomy, Johns Hopkins University, Baltimore, Maryland 21218, USA}


\date{\today}

\begin{abstract}
Theoretical description of anisotropic systems, such as layered superconductors and coupled spin chains, is often a challenge due to the different natures of interactions along different directions. As a model of such a system, we present an analytical study of $d$-dimensional ``nodes" arranged as the vertices of a Bethe lattice, where each node has nonzero spatial dimension and is described by an $O(N)$ quantum rotor model, and there is hopping between neighboring nodes. In the limit of large connectivity on the Bethe lattice, the hopping can be treated by constructing a self-consistent effective action for a single node. This procedure is akin to dynamical mean field theory, but generalized so that spatial as well as quantum fluctuations are taken into account on each node. The quantum phase transition is studied using this effective action for both infinite and finite $N$. The importance of the Perron-Frobenius uniform mode on the Bethe lattice is discussed, and its elimination via an ``infinite range hopping" term shifts the transition, leading to nontrivial critical behavior. We calculate critical exponents and find that the internode hopping reduces the upper and lower critical dimensions each by one, indicating that--at least for the purposes of quantum criticality--the large number of internode couplings is similar to adding a single extra dimension to the theory describing a single node.
\end{abstract}

% \pacs{}
%\keywords{}

\maketitle

%\section{}
%\subsection{}
%\subsubsection{}

\section{Introduction}

It is often the case that the relevent degrees of freedom in a condensed matter system are best understood via an anisotropic description, where interactions may be strong in some subset of spatial dimensions, and weak or irrelevent in others. The most obvious examples include the high temperature superconductors, where bulk theories are often constructed by studying an individual copper-oxygen plane and completely neglecting interactions between such planes.  In the recently discovered iron pnictide and chalcogenide superconductors \cite{johnston10}, such an extreme decopuling makes less sense. In this case the interlayer interactions must be taken into account, although they can generally be treated on some simpler footing than those within a plane.  Other examples include the study of dimensional reduction in organic conductors, which can be effectively described as coupled one-dimensional chains.\cite{jerome82}  While there are a substantial number of theoretical tools at one's disposal for investigating the properties of these chains or layers, treating the hopping between them simultaneously in such a way that analytical progress can be made is often a challenge.  Some insight can be gleaned from dynamical mean field theory (DMFT), where the hopping between sites in a system is treated in a self-consistent fashion, resulting in a theory of decoupled sites which retain quantum---but not spatial---fluctuations. \cite{georges96} This approach is known to become exact in the limit of large connectivity, {\em i.e.}\ for each site coupled to a large number of nearest neighbors. In general the self-consistent equations of DMFT are complicated and often must be solved numerically.  A simplification occurs, however, when the sites reside on a Bethe lattice, in which case the self-consistent equations can be solved in closed form and analytical progress can be made more readily.

Since its introduction in 1935, the Bethe lattice \cite{bethe35} has played a distinguished role in statistical mechanics and many body physics. From a theoretical standpoint, the tree-like structure of the Bethe lattice, which inhibits loops and ensures that there is a unique path between any two nodes, often leads to tremendous simplifications and in some cases makes exact solutions possible. \cite{baxter82,semerjian09} Although there is of course no physical realization of such a lattice in Nature, it can sometimes serve as a useful approximation to real world systems in certain situations.  Under some circumstances the behavior of a Bethe lattice may approximate that of a physically realizable lattice having the same connectivity. One classic example is the low-temperature expansion of the Ising model, where one finds that the results at the first two orders depend only on the number of sites and the number of nearest neighbors for each site, so that at this order the results are identical for a (hyper)cubic lattice in $d$ dimensions and for a Bethe lattice with connectivity $z=2d$. \cite{baxter82} Conversely, it is sometimes found that the Bethe lattice effectively behaves as a one-dimensional system regardless of its connectivity. Indeed, one of the main results of the work presented here is that the upper and lower critical dimensions for a strongly coupled critical theory on a Bethe lattice at large connectivity are each shifted by one relative to the theory with no Bethe lattice coupling, so that the Bethe lattice structure effectively adds one extra dimension. Further evidence for the validity of this approach comes from the recent study of layered iron-pnictide superconductors, where quantitative agreement with experimental results was obtained by approximating a superconductor---which can be thought of as a one-dimensional stack of two-dimensional layers---as a Bethe lattice of layers in the large-connectivity limit. \cite{murray10} While the underlying reason for the similar critical behavior of one-dimensional systems and Bethe lattice models appears to hinge on the fact that neither system has loops, a better understanding of this correspondence is clearly desirable and is a motivation for the present study. In general, one must always keep in mind the limitations of the Bethe lattice as a model for physical lattices and exercise caution in relating results between these two systems.

In this work we consider Bethe lattice systems in which the nodes themselves are of finite spatial extent, as in the layered superconductor example mentioned above. Throughout this paper, the term ``node" shall refer to a {\em finite-dimensional} object composed of $d$ spatial dimensions.  For example, such a model might describe a set of parallel, one-dimensional $(d=1)$ spin chains that are coupled to each other as shown in Figure \ref{cartoon}(a), or a layered system in which each two-dimensional $(d=2)$ layer is coupled to neighboring layers above and below. Such an approach has been considered previously in the context of coupled Luttinger chains \cite{arrigoni99,georges00}, but the emphasis in those studies was on the application of numerical techniques from DMFT to obtain information about the phase diagrams of the systems under consideration. As we mentioned above, the method has also been applied to coupled superconducting layers, where the resulting effective theory in two spatial dimensions was used to obtain expressions for the fluctuation conductivity and magnetization. \cite{murray10} Here we study quantum criticality in such systems under more general circumstances and using an entirely analytical approach. It is worth stressing that the approach presented here is distinct from other DMFT extensions such as ``cluster" DMFT, in which impurities having some nonzero but finite spatial extent are embedded into a lattice. In contrast, the ``impurities" in the present study are of infinite spatial extent.

We focus here on the example of the quantum rotor model because of its relative simplicity, although the main results of this work can be applied to a wide variety of quantum and classical models. In particular, many of the results below can be applied directly to the $\phi^4$ model with large $N$, which is known to share many of the same critical properties as the rotor model. \cite{zinnjustin02} The quantum rotor model is known to emerge as a low-energy effective theory for certain Heisenberg antiferromagnets ($N=3$), as well as the Bose-Hubbard model ($N=2$). \cite{sachdev11} Constraining our discussion to the quantum rotor model thus allows us to address basic questions about the effects of the internode coupling on critical exponents and the upper and lower critical dimensions in a relatively simple and general setting.  In addition, some mean field results for nodes of finite dimension can be directly compared to those found for similar models with $d=0$ using the \emph{spectral dimension} of the Bethe lattice. \cite{burioni96, cassi99, burioni00}

The remainder of this paper is organized as follows. In Section II we introduce our model of coupled quantum rotor nodes and self-consistently derive an effective action for a single node in the limit of large connectivity. The appropriate limits of this effective action for describing the ordered and disordered phases are then discussed. In Section III we study the critical point in the large $N$ limit and determine the upper and lower critical dimensions. In Section IV we consider $1/N$ corrections and derive results for the self-energy in various dimensions. Finally, in Section V we discuss our results and suggest possibilities for future work in this area.

\section{Self-consistent action in the large $z$ limit}

We begin with the action describing a quantum rotor model for an $N$-component field $\phi$ at zero temperature on a set of ``nodes" forming a Bethe lattice (or, equivalently, a large random graph with fixed connectivity $z$), where each node corresponds to a quantum critical system of spacetime dimension $d+1 \geq 1  $:
\be 
\label{action} 
S_\mathrm{graph} =  \sum_i \left( S_i + S_i^\mathrm{hop} \right) 
\ee
The first term describes a $N$-component quantum rotor on each node $i$: 
\be 
\begin{split} 
\label{action_i} 
S_i = & \frac{1}{2}\int d^{d+1}x \phi_i^a (x) \hat G_0^{-1} \phi_i^a (x) - \frac{i}{2g} \int d^{d+1}x \lambda_i (x) \\ & + \frac{i}{2} \int d^{d+1}x \lambda_i (x) \phi_i^a(x)\phi_i^a(x), 
\end{split} 
\ee 
where $g$ is a dimensionless parameter controlling the strength of quantum fluctuations, the field $\lambda$ enforces the constraint $ \phi_i^a (x) \phi_i^a(x) = 1$ at every point in space, and there is an implicit summation over $a = 1,\ldots,N$. The strength of the fluctuations of the $N$-component field $\phi (x)$ are controlled by $g$ in \eqref{action_i}. As $g$ is varied, the system is known \cite{sachdev11,chubukov94} to exhibit a quantum critical point at $g=g_c$. For $g>g_c$ the system is in a disordered phase with $\left< \phi \right> = 0$, while for $g<g_c$ the system realizes an ordered phase in which $\phi$ becomes ordered along some direction. Throughout this study we use relativistic notation to describe the zero temperature quantum system, with $p \sim (\omega, \mathbf{p})$, and where we have set a velocity equal to one. Although much of what follows will be true for a general bare propagator $G_0$, for concreteness we can take $G_0^{-1}(p) = -\partial_\mu^2 +r_0$, with $r_0 \geq 0$. (Note that $r_0$ is shifted from zero to a finite value upon including the simple tadpole correction to the propagator. It can be checked that it makes no difference whether this correction is included before or after the large $z$ limit is employed to obtain the self-consistent effective action.)

The second term in \eqref{action} describes hopping between a given node $i$ and its $z$ neighbors, denoted by $j$:
\be
\label{hop}
S_i^\mathrm{hop}= - \frac{t_0}{2} \sum_{j(i) = 1}^z \int d^{d+1}x \phi_i^a (x) \phi_j^a (x).
\ee
The goal is now to recast the internode hopping described by \eqref{hop} as a modification to the action \eqref{action_i} corresponding to a single node. Such a procedure shall be shown to be valid in the large $z$ limit. In order to describe the transition to the condensed phase (with $\left< \phi^a_i \right> \neq 0$) consistently, it is necessary to eliminate the zero mode solution on the graph, also known as the uniform Perron-Frobenius mode, which is a general spectral feature of Bethe lattices or, equivalently, large random regular graphs. This mode contributes an isolated delta function at low energy to the density of states on the graph, and in a bosonic theory there is a tendency to condense into this mode. \cite{laumann09} In order to avoid such a condensation and obtain nontrivial critical behavior, we introduce a constraint that eliminates this uniform mode from the spectrum of the graph: 
\be
\label{constraint}
\sum_i \phi_i^a (x) = 0.
\ee
This condition is enforced by introducing the field $\chi(x)$, which acts as a Lagrange multiplier. This constraint can also be thought of as an ``infinite range hopping" that has equal amplitude between every pair of nodes, as can be seen by adding a term $\sim \chi^2$ to the action, completing the square and integrating out the auxiliary field. 

Let
\be
S [\phi_i, \lambda_i, \chi] \equiv S_i[\phi_i, \lambda_i] + i z t_0 \int d^{d+1}x \chi^a(x) \phi_i^a(x)
\ee
be the action describing a single node. We shall now implement the procedure described above in a self-consistent fashion by letting $j=1,2,\ldots,z$ label the nearest neighbors of node $i=0$ and assuming that we can integrate all further neighbors out from the partition function. 
\begin{figure}
\includegraphics[width=0.4\textwidth]{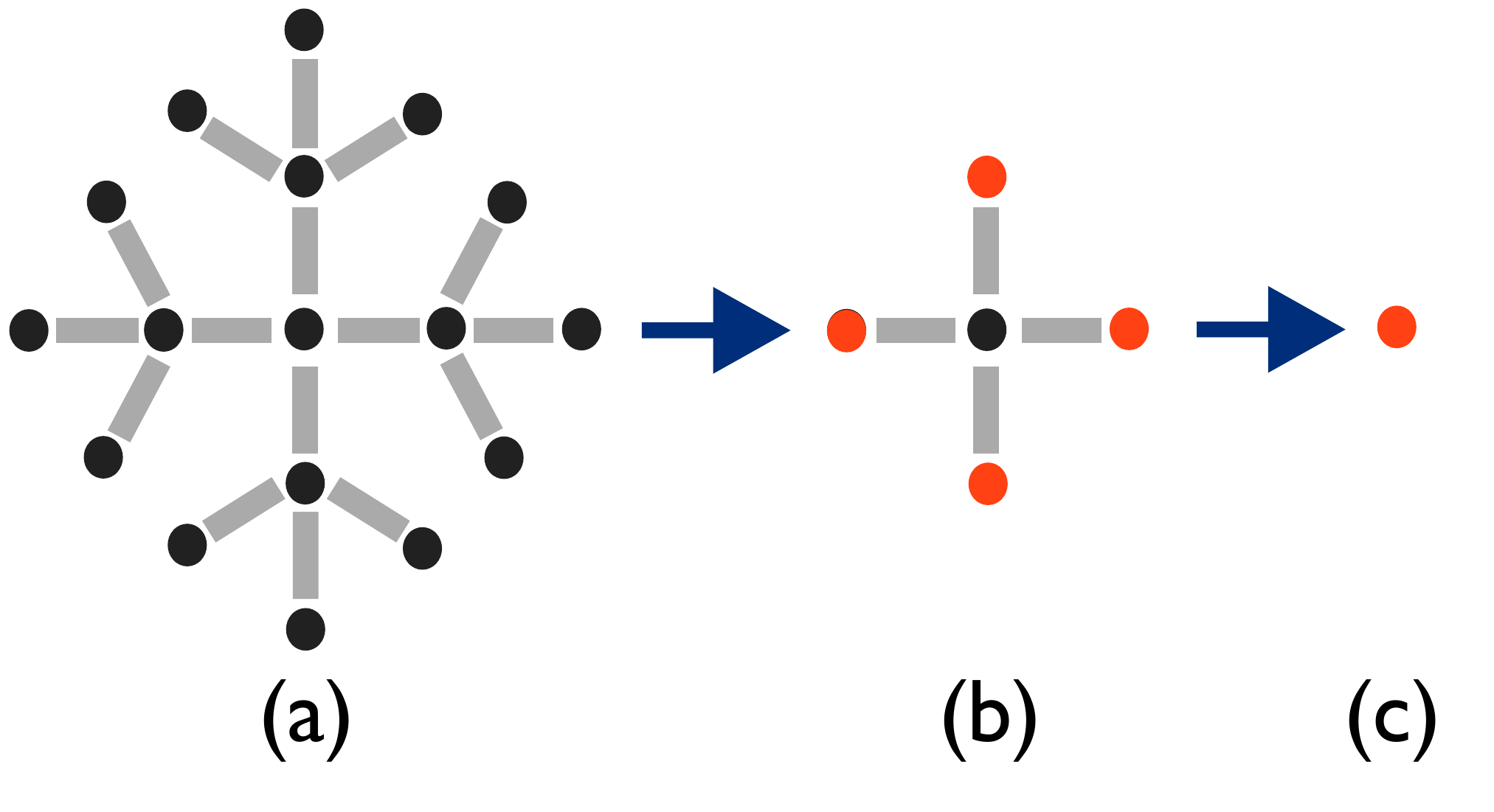}
\caption{Schematic depiction of the self-consistent calculation of the effective action \eqref{s_eff}. (a) We begin with a set of nodes on a Bethe lattice, where each node describes a $d$-dimensional system and is coupled to a large number of nearest neighbors. (b) It is assumed---to be verified self-consistently later on---that further neighbors can be integrated out, giving a correction to the action of the nodes shown in red. (c) The integration of the remaining nearest neighbors is performed explicitly, leaving a single node decoupled from its neighbors and described by the effective action \eqref{s_eff}.
\label{cartoon}}
\end{figure}
The procedure is shown schematically in Figure \ref{cartoon}. We assume---intending to self-consistently verify this assumption below---that this has the effect of modifying the action for the nodes denoted by $j$ in the large $z$ limit, {\em i.e.} $S [\phi_i, \chi] \to \tilde S [\phi_i, \chi]$. The partition function describing the $i=0$ node is then (neglecting the normalization)
\be
\label{z}
\begin{split}
Z(0) \sim & \int {\cal D} \chi {\cal D} \phi_0 {\cal D} \lambda_0 e^{- S [\phi_0, \lambda_0, \chi]}  
\\ & \times \prod_{j=1}^z  \left( \int {\cal D} \phi_j {\cal D} \lambda_j e^{- S [\phi_j, \lambda_j, \chi]} e^{t_0 \int d^{d+1}x \phi_0^a \phi_j^a} \right) .
\end{split}
\ee

The next step is to perform the functional integrals over $\phi_j$ explicitly, thus obtaining the self-consistent expression for $\tilde S [\phi_0,\chi]$. Expanding the last exponential in \eqref{z} yields
\be
\label{expand}
\sum_{n=0}^\infty \frac{t_0^n}{n!} \left[ \int d^{d+1}x \phi^a_0 (x) \sum_{j=1}^z \phi^a_j (x) \right]^n.
\ee
When the part of this expression in square brackets is multiplied out, a typical term will look like
\be
\label{hop_term}
\begin{split}
& \left( \int d^{d+1}x_1 \phi^a_0 (x_1) \phi^a_{j_1} (x_1) \right) \left( \int d^{d+1}x_2 \phi^b_0 (x_2) \phi^b_{j_2} (x_2) \right)
\\ & \times \ldots \left( \int d^{d+1}x_n \phi^c_0 (x_n) \phi^c_{j_n} (x_n) \right).
\end{split}
\ee

Now consider the limit $z \to \infty$, with $z>n$. There will be terms such as \eqref{hop_term} with no repeated indices $(j_1 \neq j_2 \neq \ldots \neq j_n)$, terms with one repeated index, {\em etc}. In general, there will be $z^{n-m}n! /(n-m)!$ terms with $m$ repeated indices. (Note that terms with a single index appearing three or more times are neglected, as these terms will be of lower order in $z$.) The sum in \eqref{expand} thus becomes 
\be
\label{hop_term2}
\begin{split}
& \sum_{n=0}^\infty \sum_{m=0}^{n/2} \frac{t_0^n z^{n-m}}{(n-m)!} 
\left( \int d^{d+1}x \phi^a_0 (x) \left< \phi^a_j (x) \right> \right) ^{n-m}
\\ & \times \left( \int d^{d+1}x \int d^{d+1} x' \phi^b_0 (x) \left< \phi^b_j (x) \phi^c_j (x') \right> \phi^c_0 (x') \right) ^m .
\end{split}
\ee
The second summation in the above expression runs to $n/2$ if $n$ is even, or to $n/2-1$ if $n$ is odd. We have also performed the functional integrals over $\phi_j$, yielding expectation values defined with respect to $\tilde S [\phi_j, \chi]$, as can be seen from \eqref{z}. The sums in \eqref{hop_term2} can be performed explicitly, resulting in the expression
\be
\label{hop_resummed}
\begin{split}
& \frac{z t_0 e^{z t_0 \int \left< \phi \right> \phi}  \int \left< \phi \right> \phi - z t_0^2 e^{ z t_0^2 \int \int \phi \left< \phi \phi \right> \phi} \int \int \phi \left< \phi \phi \right> \phi }
{z t_0 \int \phi \left< \phi \right> - z t_0^2 \int \int \phi \left< \phi \phi \right> \phi}
\\ & \equiv \exp \left( \Gamma [\phi , \chi] \right).
\end{split}
\ee
The integration variables and vector indices in this expression have been suppressed for brevity, but they can be restored by comparing with \eqref{hop_term2}, which features the same integral expressions. The node indices have also been dropped, which is permissible since the layers are indistinguishable and the only remaining field that has not been integrated out is $\phi_0 \equiv \phi$. Note in particular that $\Gamma$ is a functional of both $\phi$ and $\chi$, with the latter dependence arising indirectly through the expectation values $\left< \phi \right>$ and $\left< \phi \phi \right>$. 

We thus obtain the following result for the partition function and effective action describing node $i=0$:
\be
\label{s_eff}
\begin{split}
Z(0) &= \int {\cal D} \phi {\cal D} \chi {\cal D} \lambda e^{- \tilde S [\phi, \lambda, \chi]}
\\ \tilde S [\phi, \lambda, \chi] &= S [\phi , \lambda, \chi] - \Gamma [\phi , \chi].
\end{split}
\ee
Ideally, one would like to calculate $\left< \phi (x) \right>$ and $\left< \phi (x) \phi (x') \right>$ explicitly with respect to \eqref{s_eff} in order to obtain a fully self-consistent effective action in terms of the fields $\phi$ and $\chi$. However, the unwieldy form of $\Gamma$ makes this prohibitively difficult, so we shall have to content ourselves with various approximations. There are two especially interesting cases to consider, which describe the ordered ($\left< \phi \right> \neq 0$) and disordered ($\left< \phi \right> = 0$) phases, respectively. These limits are considered in turn in the following subsections.

\subsection{Ordered Phase}

Let us first assume that the system is ordered, which means that one component of $\phi$ has a nonzero expectation value. Letting $\phi = (\sigma, \vec \pi)$, we consider the case where $\left< \sigma \right> \neq 0$, and $\vec \pi$ describes the remaining $N-1$ components. From the form of \eqref{hop_resummed}, one sees that in this case it makes sense to rescale the hopping amplitude as $t' \equiv z t_0$, which is to remain finite as $z \to \infty$ and $t_0 \to 0$. The importance of this particular scaling to obtaining a nontrivial description of the condensed phase has been previously noted in the context of bosonic dynamical mean field theory. \cite{byczuk08,anders11} In this case the terms in \eqref{hop_resummed} with coefficients $\sim z t_0^2$ disappear in the large $z$ limit, and the effective action in \eqref{s_eff} reduces to 
\be
\label{s_ord}
\begin{split}
& \tilde S_\mathrm{ord} [ \phi, \lambda, \chi ] \equiv S [\phi, \lambda , \chi] - t' \int d^{d+1}x \left< \phi^a (x) \right> \phi^a (x)
\\ & = \int d^{d+1}x \bigg\{ \frac{1}{2} \sigma \hat G_0^{-1} \sigma + \frac{1}{2} \pi^a \hat G_0^{-1} \pi^a - \frac{i}{2g} \lambda
\\ & \quad + \frac{i}{2} \lambda \left( \sigma^2 + \pi^a \pi^a \right) + t' \left[ i \chi^0 - \left< \sigma \right> \right] \sigma + i t' \chi^a \pi^a \bigg\} ,
\end{split}
\ee 
Note that $\chi$ cannot be trivially integrated out of \eqref{s_ord}, since the expectation value  $\left< \sigma \right>$ depends on it. Assuming a uniform solution, the mean field equations for $\sigma$ and $\chi$ following from \eqref{s_ord} are, respectively,
\be
\label{mean_field}
\begin{split}
0 &= (r-t') \left< \sigma \right> + i t' \left< \chi^0 \right>,
\\ 0 &= t' \left< \sigma \right> \left( i - \frac{\delta \left< \sigma \right>}{\delta \chi^0} \right),
\end{split}
\ee
where $r = r_0+i \lambda |_{N=\infty}$ is the mass term for the field $\sigma$, including the saddle point value of the field $\lambda$, which we determine self-consistently in the next section. These equations have the two possible solutions
\be
\label{transition}
\begin{cases}
\left< \sigma \right> = 0, 	& r > 0 \\
\left< \sigma \right> = i \chi^0 , 	& r = 0 .
\end{cases}
\ee
This result indicates that there is an instability to an ordered phase at $r=0$. Substituting this mean field result back into the action \eqref{s_ord}, one obtains the usual action for the quantum rotor model, which is known to exhibit a second order phase transition at $r=0$. \cite{sachdev11,chubukov94}

Some additional remarks regarding this result are in order. First, it can easily be checked that if the constraint \eqref{constraint} eliminating the zero mode is not enforced, then the transition to the ordered phase occurs at $r=t'$. (This can be seen from the mean field equation for $\phi$ following from \eqref{s_ord} with the term containing $\chi (p)$ set to zero.) This would be undesirable because, as we shall see below, when studying the disordered phase it is natural to scale the hopping as $\sqrt{z} t_0 \equiv t = finite$ for $z \to \infty$; but in this case the transition would occur at $r=\sqrt{z} t$, which is infinite under this scaling, implying that the system is always in the ordered phase when the zero mode is not eliminated. Second, \eqref{transition} was obtained with the first scaling, in which $t'$ is equal to a finite constant. When approaching the transition instead from the disordered side and employing the scaling $t = finite$, it shall be seen that the transition occurs not at $r=0$, but at $r_c=2t$. This is not inconsistent, however, since the latter quantity is zero in the scaling limit that was used to obtain \eqref{transition}. We therefore see that the overall effect of the elimination of the zero mode is to shift the transition to the ordered state from $r=\infty$ to $r\sim{\cal O}(t)$.

\subsection{Disordered phase}

Now let us consider another limit of \eqref{s_eff}, in which we assume $\left< \phi \right> = 0$ and scale the hopping according to $t \equiv \sqrt{z} t_0$, which remains finite as $z \to \infty$ and $t_0 \to 0$. In this case, the effective action reduces to
\be
\label{s_dis}
\begin{split}
& \tilde S_\mathrm{dis} [ \phi, \lambda, \chi ] \equiv S [\phi , \lambda , \chi]
\\ & - \frac{t^2}{2} \int d^{d+1}x \int d^{d+1}x' \phi^a (x) \left< \phi^a (x) \phi^b (x') \right> \phi^b (x') .
\end{split}
\ee
The mean field equation for $\phi$ following from this action is
\be
\begin{split}
& 0 =  i \left< \chi^a (x) \right>
\\ & + \int d^{d+1}x' \left[ \delta(x-x') r - t^2 \left< \phi^a (x) \phi^b (x') \right> \right] \left< \phi^b (x') \right>.
\end{split}
\ee
Because in the disordered phase $\left< \phi^b (x') \right> = 0$ by assumption, this equation implies that $\left< \chi^a (x) \right> = 0$. This result is reasonable, since it effectively says that---at least at mean field level---the constraint eliminating the zero mode has no effect in the disordered phase. It is also easy to see that the same result will hold in a general interacting theory, such as a theory with a $\phi^4$ coupling. Quite generally, one sees that $\left< \phi \right> = 0 \implies \left< \chi \right> = 0$.

\section{Critical point in $N \to \infty$ limit}

With the results of the preceding section in place, we can now proceed to characterize the quantum phase transition at $g=g_c$, starting from the disordered phase in the large $N$ limit. As we saw in the last section, obtaining nontrivial behavior requires that we employ the scaling in which $\sqrt{z}t_0 \equiv t$ takes a finite value. After rescaling $g \to g/N$, this effective action \eqref{s_dis} becomes
\be
\label{action_eff}
S_\mathrm{eff} [\phi, \lambda] & = \frac{1}{2}  \int d^{d+1}x \bigg[ \phi^a \hat {\cal G}_0^{-1} \phi^a - \frac{i N}{ g} \lambda + i  \lambda \phi^a \phi^a \bigg] .
\ee
The new propagator is related to the original one by the condition
\be
\label{propagators}
{\cal G}_0^{-1}(p) \equiv G_0^{-1}(p) - t^2 {\cal G}(p),
\ee
where 
\be
\label{g_full}
{\cal G}(x-x') = \left< \phi^a(x) \phi^a(x') \right>
\ee
is the full propagator, defined with respect to the effective action \eqref{action_eff}. In writing \eqref{action_eff}, we have used the fact that $\left< \phi^a \phi^b \right> \sim \delta^{ab}$. [Note that there is no sum on $a$ in \eqref{g_full}.] Together, \eqref{action_eff}--\eqref{g_full} form a closed set of self-consistent equations describing a single node on the Bethe lattice, where the effects of hopping between nodes are included in the self-consistent propagator ${\cal G}_0$. 

In the $N \to \infty$ limit, we have ${\cal G}(p) \to {\cal G}_0(p)$, and \eqref{propagators} can be solved, yielding
\be
\label{g00}
{\cal G}_0 = \frac{2}{ p^2+r + \sqrt{(p^2+r)^2-4t^2}}.
\ee
Here we have used for the bare propagator $G_0^{-1}(p) = p^2+r$, which includes the saddle point contribution from the field $\lambda$. Although at large momenta we have ${\cal G}_0 \to G_0$, it is apparent from \eqref{g00} that there is no value of $r$ for which the propagator diverges at $p=0$. Similar non-divergent behavior has been noted in DMFT studies. \cite{georges96} The lack of a diverging susceptibility in this case can be traced to the elimination of the uniform Perron-Frobenius mode on the Bethe lattice. Such a divergence would correspond to the onset of uniform order along a given direction, but such order is forbidden by the constraint \eqref{constraint}. Rather than condensing into a state with uniform order on every node of the Bethe lattice, the system condenses into the next-lowest energy mode available, which is nonuniform from node to node and is specified by some complicated ``wave vector" on the Bethe lattice. A parallel can be made here to antiferromagnetic systems on an ordinary, say square, lattice, where the uniform ({\em i.e.}\ ferromagnetic) susceptibility remains finite due to the fact that the system does not have an instability toward uniform order, but rather a tendency to form nonuniform ordered state characterized by some nonzero wave vector. In the self-consistent large-$z$ calculation presented here, the Green's function \eqref{g00} plays in effect the role of the uniform susceptibility, and the fact that it does not diverge simply means that the lowest energy mode on the graph is nonuniform by construction.

Although there is no divergence in the Green's function, one can see that there are gapless excitations at the critical point by considering the spectral function. Returning for the moment to nonrelativistic notation and analytically continuing to real frequency, the spectral function is given by
\be
\label{spectral}
\begin{split}
{\cal A} (\omega, \mathbf{k}) =& -\frac{1}{\pi} \mathrm{Im}\ {\cal G}_0 (\omega, \mathbf{k})
\\ =& \frac{1}{2\pi t^2}\Theta \left[ \omega^2 - (\mathbf{k}^2 + r-r_c) \right] 
\\ & \quad \quad \times \sqrt{(-\omega^2 + \mathbf{k}^2 + r)^2 - r_c^2}.
\end{split}
\ee
One sees from this expression that there is a continuum of excitations, with the energy gap $\Delta = \sqrt{r-r_c}$. This continuum of states can be explained by thinking of the node described by the effective action \eqref{action_eff} as an impurity coupled to an effective medium with which the impurity can exchange energy. Such a picture is often employed in DMFT, in which case the impurity is a single quantum site.

In order to determine the value of $r$, which depends on the coupling $g$, we first integrate out the field $\phi$ from the action \eqref{action_eff}, leading to
\be
S_\mathrm{eff} [\lambda] = \frac{N}{2} \left[ \mathrm{tr\ ln}\left( \hat {\cal G}_0^{-1} + i \lambda \right) - \frac{i}{g} \int d^{d+1}x \lambda \right].
\ee
In the $N \to \infty$ limit, assuming a constant solution $i \lambda(x) = r-r_0$ and using the expression \eqref{g00} for the bare propagator, one obtains the saddle point equation
\be
\label{saddle}
\begin{split}
\frac{1}{g} &= \intp \frac{\partial}{\partial r} \ln {\cal G}_0 (p)^{-1}
\\ & = \intp \frac{1}{\sqrt{(p^2+r)^2-4t^2}}.
\end{split}
\ee

In order to examine the critical properties of the system, we consider the $p \to 0$ limit of the integrand in \eqref{saddle}. Then the integrand clearly diverges as $r \to r_c \equiv 2t$ from above, with the critical value of $g$ determined by the condition
\be
\label{g_c}
\frac{1}{g_c} = \int^\Lambda \frac{d^{d+1}p}{(2 \pi)^{d+1}} \frac{1}{\sqrt{p^4 + 4tp^2}}
\ee 
Noting that the integrand $\sim 1/p$ as $p$ approaches zero, we see that the integral is IR convergent and that the saddle point equation has a solution for $d>0$. This establishes $d=0$ as the lower critical dimension of the theory.

Taking the energy gap $\Delta = \sqrt{\delta}$, where $\delta = r - r_c$, to be small, we can investigate how it depends on the distance from the critical point at $g=g_c$. From \eqref{saddle} and \eqref{g_c}, we obtain
\be
\label{saddle_diff}
\begin{split}
\frac{1}{g_c} - \frac{1}{g} = & \intp \bigg[ \frac{1}{\sqrt{p^4 + 2 r_c p^2}} 
\\ & -  \frac{1}{\sqrt{p^4 + 2 r_c p^2 + 2 \delta (r_c+p^2) + \delta^2}} \bigg] .
\end{split}
\ee
We wish to determine the dependence of the above integral on $\delta$ for spatial dimensions $d=1,2,3$. The integral can be calculated analytically and expanded for small $\delta$. Defining
\be
\Delta \sim (g - g_c)^{z \nu},
\ee
where $z$ and $\nu$ are critical exponents, we obtain the following results:
\be
g-g_c \sim
\begin{cases}
\Delta, 	& d=1 \\
\Delta^2 \ln \Delta, 	& d=2 \\
\Delta^2,		& d=3.
\end{cases}
\ee
Due to the Lorentz invariance of our theory, the dynamical critical exponent is given by $z=1$, which remains true at all orders in $1/N$. The first result therefore corresponds to $\nu = 1$ in $d=1$. The last line corresponds to $\nu = \frac{1}{2}$ in $d=3$, which is equal to the usual mean field value for a system with no internode hopping. Since one obtains mean field critical exponents above two spatial dimensions, we thus establish $d=2$ as the upper critical dimension in our theory. Combining this result with the discussion below \eqref{g_c}, we see that, compared with the standard quantum critical theory without internode hopping, the upper and lower critical dimensions have both been shifted down by one, with the upper critical dimension shifting from $d=3$ to 2 and the lower critical dimension shifting from $d=1$ to 0. This seems to indicate that, at least for the purposes of criticality, the large number of internode couplings effectively act as one extra dimension in the theory.

\section{Corrections at Finite N}

We next evaluate the leading $1/N$ contribution to the propagator \eqref{g_full}. In order to obtain the propagator for the field $\lambda$, source terms can be inserted into the original action \eqref{action_eff}, then integrating out $\phi$ from \eqref{action_eff} and differentiating with respect to the source terms yields the propagator. This procedure leads to the effective action
\be
\begin{split}
\label{s}
S = & \frac{1}{2}  \intp \phi^a (p) {\cal G}_0^{-1}(p) \phi^a (-p) 
\\ & + \frac{N}{4} \intp \lambda (p) \Pi (p) \lambda (-p)
\\ & + \frac{i}{2} \intp \intq \phi^a (p) \lambda (q-p) \phi^a (-q).
\end{split}
\ee
The inverse propagator for the $\lambda$ field is proportional to the polarization operator:
\be
\label{pi}
\Pi (p) = \intk {\cal G}_0 (p+k) {\cal G}_0(k).
\ee
\begin{figure}
\includegraphics[width=0.4\textwidth]{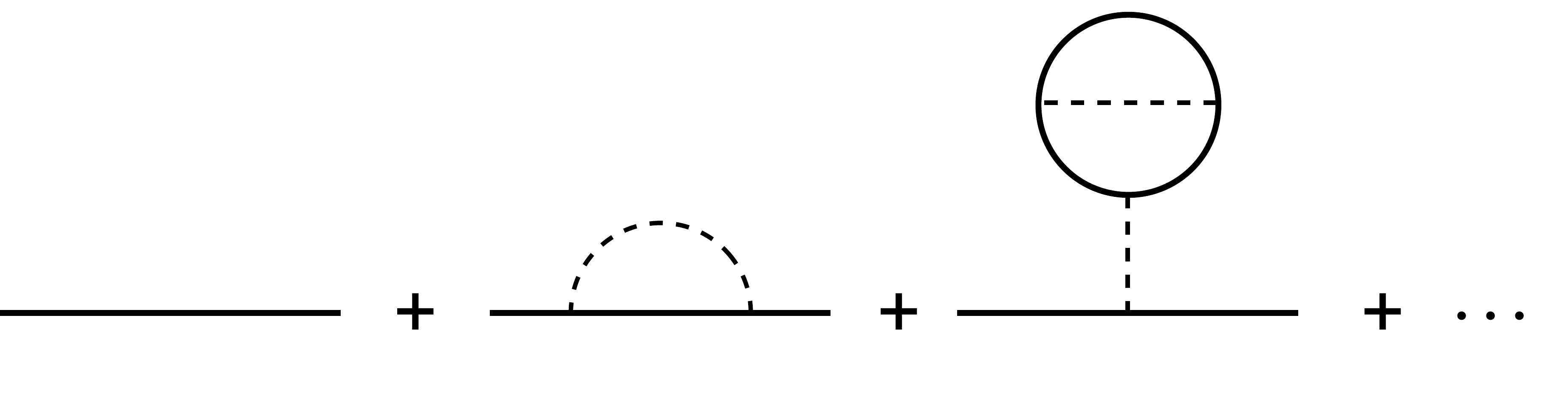}
\caption{Diagrams contributing to the propagator ${\cal G}$ up to ${\cal O}(1/N)$. The solid line represents ${\cal G}_0$, and the dashed line is the propagator for $\lambda$.
\label{diagrams}}
\end{figure}
\indent We now wish to calculate the full propagator ${\cal G}(p)$ with respect to the action \eqref{s} to ${\cal O}(1/N)$, which is shown diagrammatically in Figure \ref{diagrams} and related to ${\cal G}_0(p)$ by \eqref{propagators}. We therefore have
\be
\label{g_series}
{\cal G}(p) = {\cal G}_0(p) - {\cal G}_0(p) \Sigma (p){\cal G}_0(p) + \ldots,
\ee 
where the self-energy $\Sigma (p)$ contains the corrections from the second and third diagrams in Figure \ref{diagrams}: 
\be
\label{sigma0}
\Sigma (p) = \tilde \Sigma (p) - \frac{1}{\Pi (0)} \intq {\cal G}_0^2(q) \tilde \Sigma (q),
\ee
where
\be
\label{sigma1}
\tilde \Sigma(p) = \frac{2}{N } \intq \frac{1}{\Pi (q)} \left[ {\cal G}_0(p+q) - {\cal G}_0(q) \right].
\ee

Before explicitly calculating the self-energy, let us determine the form of the full Green's function to ${\cal O}(1/N)$, as given in \eqref{g_full}, subject to the relation between ${\cal G}$ and ${\cal G}_0$ from the self-consistency condition \eqref{propagators}. Inverting \eqref{g_series} gives
\be
{\cal G}^{-1}(p) = {\cal G}_0^{-1}(p) + \Sigma (p) + \ldots,
\ee
and combining this with \eqref{propagators} yields
\be
{\cal G}^{-1}(p) \approx G_0^{-1}(p) + \Sigma (p) - t^2 {\cal G}(p).  
\ee
This can be solved to obtain a self-consistent expression for the full propagator:
\be
\label{prop}
{\cal G}(p) \approx \frac{2}{ p^2+r +\Sigma(p) + \sqrt{[p^2+r+\Sigma(p)]^2-4t^2}},
\ee
which is valid to ${\cal O}(1/N)$. As in the $N=\infty$ case, we see that the Green's function does not diverge for any value of $r$, which again is a consequence of our elimination of the uniform mode on the Bethe lattice.

We now turn our attention to the self-energy, given in \eqref{sigma0} and \eqref{sigma1}. Since the last term in \eqref{sigma0} is momentum-independent, it simply gives a correction to the value of $r$. Such a shift does not affect the critical properties of the system, however, and so we focus on the momentum-dependent part of the self-energy, $\tilde \Sigma (p)$. Furthermore, this shift in $r$ can be omitted in the following calculation of $\tilde \Sigma (p)$, since this ${\cal O}(1/N)$ shift would lead to a contribution at ${\cal O}(1/N^2)$ in \eqref{sigma1}.

In order to calculate the self-energy, the polarization bubble $\Pi (q)$ must first be evaluated. Using \eqref{g00}, \eqref{pi} becomes
\be
\begin{split}
\label{pi2}
\Pi (q,r) = & \frac{(2t)^{(d+1)/2}}{t^2} \int \frac{d^{d+1} \bar k}{(2 \pi)^{d+1}} 
\\ & \quad \quad \left[ \bar k ^2 + \bar r - \sqrt{(\bar k ^2 + \bar r )^2 - 1} \right]
\\ & \times \left[ (\bar k + \bar q)^2 + \bar r - \sqrt{\left[ (\bar k + \bar q)^2 + \bar r \right]^2 - 1} \right] ,
\end{split}
\ee
where we have redefined the momenta as $k = \bar k \sqrt{2t}$ and $q = \bar q \sqrt{2t}$, and rescaled $r = 2t \bar r$. Careful examination shows that \eqref{sigma1} will be dominated by large-$q$ behavior, so we consider $\Pi (q)$ in this limit. Although the integral in \eqref{pi2} cannot be evaluated analytically, we can obtain approximate expressions in various dimensions by using the expansions
\be
\label{integrand}
\begin{split}
& (\bar k + \bar q)^2 + \bar r - \sqrt{\left[ (\bar k + \bar q)^2 + \bar r \right]^2 - 1} = 
\\ & \begin{cases}
\frac{1}{2 \bar q^2} + \left[ \left( \frac{4}{d+1} - 1 \right) \bar k^2 - \bar r \right] \frac{1}{2 \bar q^4} + \ldots, 	& \bar q^2 \gg \bar k^2 \\
\frac{1}{2 \bar k^2} + \left[ \left( \frac{4}{d+1} - 1 \right) \bar q^2 - \bar r \right] \frac{1}{2 \bar k^4} + \ldots, 	& \bar k^2 \gg \bar q^2 .
\end{cases}
\end{split}
\ee
Here we have neglected terms that integrate to zero and replaced $(\bar q \cdot \bar k)^2/(\bar q^2 \bar k^2) \to 1/(d+1)$, as appropriate when averaging over angular integrals. 
Careful analysis of the integral in \eqref{pi2} for both large and small $k$ reveals the following behaviors of $\Pi (q)$ at large $q$:
\be
\label{pi3}
\Pi (q \to \infty ) \sim
\begin{cases}
\frac{1}{\sqrt{t} q^2}, 			& d=0 \\
\frac{\ln (q / \sqrt{2 t})}{q^2},	& d=1 \\
\frac{1}{q},					& d=2.
\end{cases}
\ee
In obtaining these expressions, we have again used $\bar q \equiv q / \sqrt{2 t}$. Numerical factors have been dropped since our only goal thus far has been to obtain the asymptotic behavior of $q$ in the large $q$ limit. By comparing with \eqref{sigma1}, it can be seen that $\tilde \Sigma (p)$ will be UV divergent in spatial dimensions $d=1,2$, but finite in $d=0$.

\begin{figure}
\includegraphics[width=0.4\textwidth]{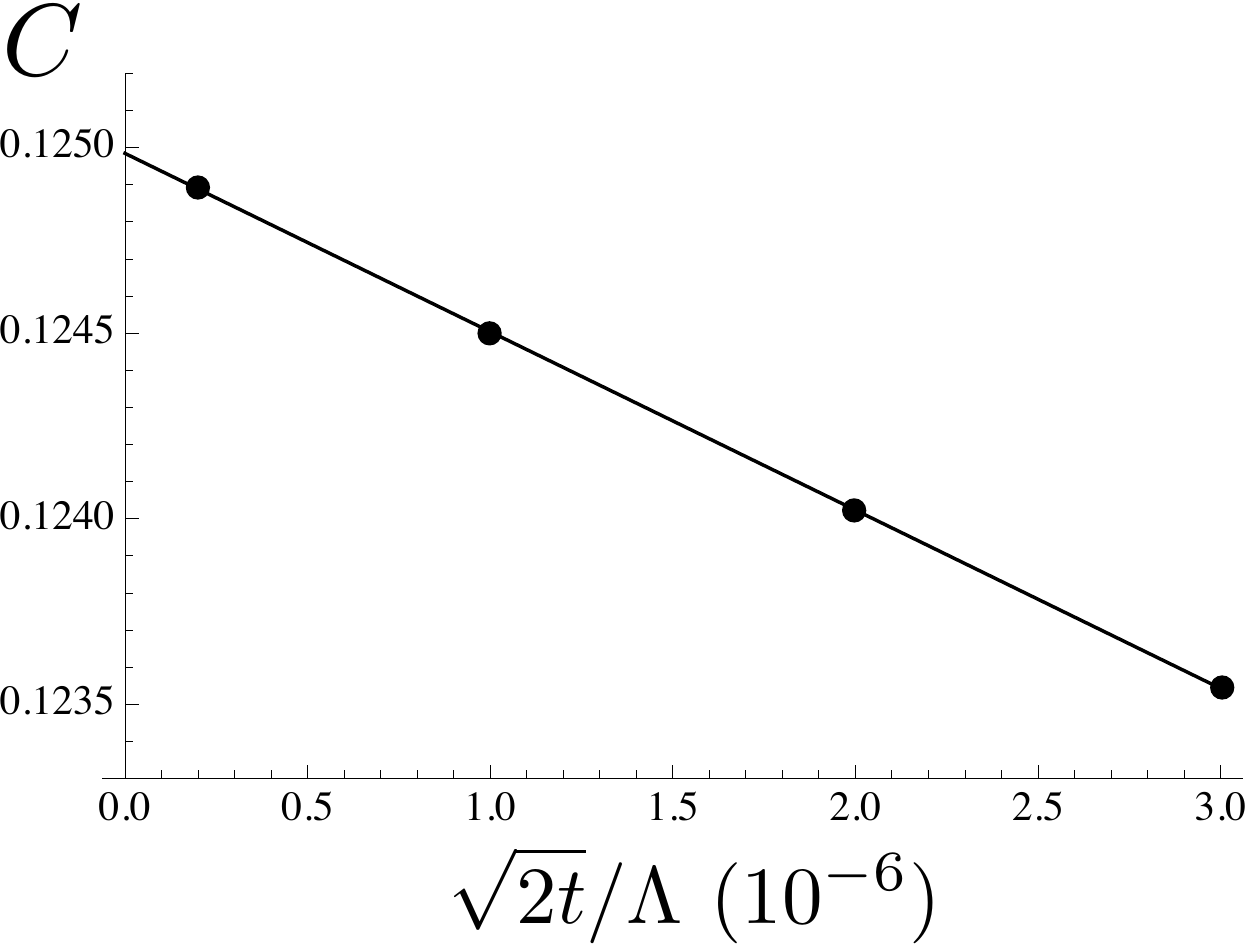}
\caption{Values of the coefficient of $\Pi (q \to \infty)$ from \eqref{pi_2d} for several values of the ultraviolet cutoff $\Lambda$. \label{alpha}. Error bars are roughly the size of the points in the figure.}
\end{figure}

With this information in hand, we can obtain a more precise result by performing the integral in \eqref{pi2} numerically for several large values of $\bar q$, then fitting the result to the appropriate form from \eqref{pi3}.  Due to the weak dependence of the expansion \eqref{integrand} on $r$, it is found that the precise value of $r$ doesn't affect the large $q$ behavior as long as $r \ll \Lambda^2$. We therefore set $\bar r = 1$ in the integrand and proceed to calculate \eqref{pi2} numerically. Specializing to $d=2$ spatial dimensions, the integral is performed for many values of $\bar q \gg 1$, and the results are fitted to the expression
\be
\label{pi_2d}
\Pi (q) \approx \frac{C}{q},
\ee
where $C$ is a fitting parameter. This is done for several values of the upper cutoff $\Lambda$, and the resulting values of $C$ are shown in Figure \ref{alpha}.
Extrapolating to the $\Lambda = \infty$ limit, the result is seen to be consistent within our numerical precision with $C = 1/8$, the standard result for a $d=2$ system with no internode hopping.

With these results, we can proceed to calculate the self-energy. Using \eqref{pi_2d}, we have (still in $d=2$)
\be
\label{sigma2}
\begin{split}
\tilde \Sigma(p) & \approx \frac{4t}{N}
\int \frac{d^{3}\bar q}{(2 \pi)^{3}} \frac{2 \bar q}{C}
\bigg[ (\bar p + \bar q)^2 - \bar q^2
\\ & \quad - \sqrt{\left[ (\bar p + \bar q)^2 + \bar r \right]^2 - 1} +\sqrt{\left( \bar q^2 + \bar r \right)^2 - 1}  \bigg] 
\\ & \approx \frac{4 t}{C N} \int \frac{d^{3}\bar q}{(2 \pi)^{3}} \frac{\bar p^2}{3 \bar q^3}
\\ & = \frac{p^2 \ln ( \Lambda / p )}{3 \pi^2 C N},
\end{split}
\ee
where the expansion \eqref{integrand} was again used in obtaining the second line. If $C = \frac{1}{8}$ from \eqref{pi_2d}, this agrees exactly with the standard result for $d=2$ with no internode hopping. This correspondence is again related to the fact that $\bar r$ does not appear at leading order in the expansion of the integrand in \eqref{sigma2}. In the standard theory, one would interpret the coefficient of the momentum-dependent part of the self-energy as the anomalous dimension $\eta$, since one could express the propagator as $G(p) \sim 1/p^{2-\eta}$. One sees from the form of \eqref{prop}, however, that, while this interpretation applies at momenta $p^2 \gg r$, for small momenta one cannot define an anomalous dimension due to the fact that the Green's function remains finite rather than diverging as a power law as $p \to 0$.

Turning briefly to the $d=1$ case, we can use \eqref{pi3} and follow a similar procedure to obtain the self-energy 
\be
\tilde \Sigma (p) \approx \frac{2 p^2}{N} \left[ \ln \left( \ln \frac{\Lambda}{\sqrt{2t}} \right) -  \ln \left( \ln \frac{p}{\sqrt{2t}} \right) \right].
\ee
This double log divergence is also found in the standard $d=1$ case with no internode hopping. \cite{polyakov87} While it does contribute a (very mildly) diverging term to the self energy, it does not lead to an anomalous critical exponent in the usual sense. 

To summarize, in this section we have derived the self-consistent expression \eqref{prop} for the full propagator including self-energy corrections. The self-energies in $d=1,2$ dimensions were then computed at leading order in $1/N$ and were found to have the same form as the standard results from the case where there is no internode hopping. This is especially interesting in light of the results of Section III, where it was shown that various quantities including critical exponents, the upper and lower critical dimensions and the position of the quantum critical point all depend on the internode hopping.

\section{Discussion}

We have presented a self-consistent, analytical treatment of coupled nodes of spatial dimension $d \geq 0$ on a Bethe lattice. We have derived and studied an effective action \eqref{s_eff} which recasts the system as an ensemble of decoupled nodes, allowing one to investigate the critical properties on both sides of the quantum phase transition. We found that there is a nontrivial quantum critical point once the zero mode has been eliminated and the internode hopping is scaled appropriately. Although there are gapless excitations at the critical point, there is no divergence in the susceptibility, which is a consequence of the suppression of the zero energy mode. A calculation of the critical exponents at the transition indicates that the upper and lower critical dimensions are shifted down by one, to two and zero spatial dimensions, respectively. This manifestation of the large connectivity limit is in contrast to the expectation from the spectral dimension of the Bethe lattice that would predict $d=0$ to be the upper critical dimension \cite{burioni00}. It is interesting that, despite this change, the self-energy appears not to be affected by the internode hoppings. This seems to be due to the fact that, while the calculation of $\nu$ involves only behavior at small momenta, the calculation of the self-energy involves integrals that are ultraviolet divergent, and these divergences are unaffected by the internode hoppings as long as $\Lambda \gg t$. 

One can imagine many possible extensions to the work presented here. For example, one could investigate how these results might be different for internode hopping along other, more realistic lattices rather than the Bethe lattice. This may require some numerical work to evaluate the self-consistent equations [which cannot in general be expressed in closed form as in \eqref{propagators}], but it would be interesting to see the extent to which results that we have derived for the highly idealized Bethe lattice would carry over to physically realizable systems. It would also be desirable to generalize to finite temperature, and to extend our results to fermionic systems, in which case contact could be made with the many results from DMFT relating to the fermion Hubbard model and related models. Investigating the full effective action \eqref{s_eff} numerically for finite $z$ may allow one to obtain a more complete description of the quantum phase transition. One could also numerically study the problem of coupled extended nodes on a Bethe lattice (or, equivalently for most purposes, a large regular random graph) directly for finite $z$ and compare the results with those derived from the self-consistent, infinite-$z$ theory presented in this work. \cite{delmaestro11}

\begin{acknowledgments}
We would like to thank Predrag Nikolic for helpful discussions. This work was supported by the Johns Hopkins-Princeton Institute for Quantum Matter, under No. DE-FG02-08ER46544 by the U.S. Department of Energy, Office of Basic Energy Sciences, Division of Materials Sciences and Engineering.
\end{acknowledgments}

%reference

\bibliographystyle{apsrev}

\begin{thebibliography}{10}

\bibitem{johnston10} D. C. Johnston, Adv. Phys. {\bf 59}, 803 (2010).

\bibitem{jerome82} D. J\'{e}rome and H. J. Schulz, Adv. Phys. {\bf 31}, 299 (1982).

\bibitem{georges96} A. Georges, G. Kotliar, W. Krauth and M. J. Rozenberg, Rev. Mod. Phys. {\bf 68}, 13 (1996).

\bibitem{bethe35} H. A. Bethe, Proc. Roy. Soc. London A {\bf 150}, 552 (1935).

\bibitem{baxter82} R. J. Baxter, \textit{Exactly Solved Models in Statistical Mechanics},  Academic Press (1982).

\bibitem{semerjian09} G. Semerjian, M. Tarzia and F. Zamponi, Phys. Rev. B {\bf 80}, 014524 (2009).

\bibitem{murray10} J. M. Murray and Z. Te\v{s}anovi\'{c}, Phys. Rev. Lett. {\bf 105}, 037006 (2010).

\bibitem{arrigoni99} E. Arrigoni, Phys. Rev. Lett. {\bf 83}, 128 (1999); E. Arrigoni, Phys. Rev. B {\bf 61}, 7909 (2000).

\bibitem{georges00} A. Georges, T. Giamarchi and N. Sandler, Phys. Rev. B {\bf 61}, 16393 (2000).

\bibitem{zinnjustin02} J. Zinn-Justin, \textit{Quantum Field Theory and Critical Phenomena }, Claredon Press (2002).

\bibitem{sachdev11} S. Sachdev, \textit{Quantum Phase Transitions}, Cambridge Univ. Press (2011).

\bibitem{burioni00} R. Burioni, D. Cassi, and C. Destri, Phys. Rev. Lett. \textbf{85}, 1496 (2000).

\bibitem{cassi99} D. Cassi and L. Fabbian, J. Phys. A: Math. and Gen. \textbf{32}, L93 (1999).

\bibitem{burioni96} R. Burioni and D. Cassi, Phys. Rev. Lett. \textbf{76}, 1091 (1996).

\bibitem{chubukov94} A. V. Chubukov, S. Sachdev and J. Ye, Phys. Rev. B {\bf 49}, 11919 (1994).

\bibitem{laumann09} C. R. Laumann, S. A. Parameswaran and S. L. Sondhi, Phys. Rev. B {\bf 80} 144415 (2009).

\bibitem{byczuk08} K. Byczuk and D. Vollhardt, Phys. Rev. B {\bf 77}, 235106 (2008).

\bibitem{anders11} P. Anders, E. Gull, L. Pollet, M. Troyer and P. Werner, New J. Phys. {\bf 13}, 075013 (2011).

\bibitem{polyakov87} A. M. Polyakov, \textit{Gauge Fields and Strings}, CRC Press (1987).

\bibitem{delmaestro11} A. Del Maestro, J. M. Murray and Z. Te\v{s}anovi\'{c}, \textit{unpublished}. 

\end{thebibliography}

\end {document}